# Broadband resonant calibration-free complex permittivity retrieval of liquid solutions

Dmitry S. Filonov[1,2,=], Egor I. Kretov[1,=], Sergei A. Kurdjumov[1], Viacheslav A. Ivanov[1] and Pavel Ginzburg[2]

*1. Department of Nanophotonics and Metamaterials, ITMO University, St-Petersburg, Russia, 197101*
*2. School of Electrical Engineering, Tel-Aviv University, Tel-Aviv, Israel, 69978*
=contributed equally

Material susceptibilities govern interactions between electromagnetic waves and matter. They are of a crucial importance for basic understanding of natural phenomena and for tailoring practical applications. Here we present a new calibration-free method for relative complex permittivity retrieval, which allows use of inexpensive and accessible equipment, and simplifies the measurement process. This method combines the advantages of resonant and non-resonant techniques, which allow the extraction of material parameters of liquids and solids in a broad frequency range, where material's loss tangent is less than 0.5. The essence of the method is based on excitation of magnetic dipole resonance in a spherical sample with variable dimensions. The size-dependent resonant frequencies and the quality factors of magnetic dipolar modes are mapped on real and imaginary parts of permittivity by employing Mie theory. Samples are comprised of liquid solutions, enclosed in stretchable covers, which allows changing the dimensions continuously. This approach enables the tuning of the magnetic dipolar resonance over a wide frequency range, effectively making resonance retrieval method broadband. The technique can be extended to powders and solid materials, depending on their physical parameters, such as granularity and processability.

## I. INTRODUCTION

The retrieval of material parameters is of a crucial importance for a broad range of applications, based on propagation of electromagnetic waves and their interactions with matter[1]. Material`s susceptibilities govern constitutive relations, substituting Maxwell's equations and enabling to find their unique solutions. While a general concept of the parametric retrieval is quite universe and is always based on probing a structure's response to an applied electromagnetic field, practical constrains set numerous technical limitations and give rise to development of many complementary experimental techniques.

The set of experimental methods for parametric retrieval is rather large, but the vast majority can be divided into two main groups: resonant [2] and non-resonant [3]. Each group has its own advantages and drawbacks: while non-resonant methods are naturally broadband, resonant approaches are restricted to a limited frequency range with an advantage to provide more accurate data. The main approaches will be briefly surveyed, and the limitations highlighted. Regardless of the type of methods, most of them require the use of a vector network analyzer (VNA) to perform measurements of the reflection parameters at a certain frequency range. In addition, specialized equipment for sample positioning or shaping is also needed. These requirements are related to some technical constrains in methods' implementation.

*Non-resonant methods*
*Open-ended coaxial probe*
An open-ended coaxial probe technique [4] is commonly used for quality control [5], antenna design [6] and in biomedical applications[7]. Specially designed probes are immersed in a liquid or placed on top of a solid. The $S_{11}$ parameter (complex-valued reflection) is then measured. The accuracy of the method is constrained by need for three-step calibration process. Interpretation of the results requires non-trivial processing, and calls for rather sophisticated software.

*Rectangular waveguide*
Another widely used method for parametric retrieval utilizes a waveguide configuration. Here a sample is inserted into a two-port waveguide section connected to a VNA. The retrieval is based on relating S-parameters to the impedance of an unknown material[8,9]. This approach allows extracting both permittivity and permeability of a dielectric slab by employing relatively straightforward mathematical formulation[10,11]. The commonly used Nicolson-Ross-Weir method is appropriate for solid samples thicker than half of a wavelength. Waveguide geometry has significant limitations when applied to liquid solutions, however cuvettes can be designed to hold liquids. Solid samples also require special shaping in order to fit waveguide cross-sections. Furthermore, the single mode operation of a waveguide limits the operational bandwidth provided by a single device. The use of several waveguides (one per each frequency band) is possible, but ends up by relatively high investments in relevant equipment. Also, a number of calibrated cuvettes for liquid solutions will be required.

*Transmission lines*
Another widely used approach utilizes transmission line geometry. A dielectric sample is placed on top of a microstrip line, connected to VNA ports. S-parameters are then converted to complex permittivity of the sample [12]. This approach allows extraction over a wide frequency range (~0.05-75 GHz) and is also applicable to lossy materials. Accurate retrieval requires a VNA, which may be expensive, in the case of high frequency measurements. The investigation of liquid solutions also requires the use of cuvettes.

*Resonant Methods*
*Resonant cavity*
Cavity configuration is one of the most commonly used resonant methods. It allows the extraction of permittivity values from resonant responses, measured with and without a sample, incorporated inside the cavity [13,14]. Here, the permittivity can be extracted only at a single (resonant) frequency. It is also worth noting that there are several extensions, where adjustable resonant cavities were reported [15,16]. Physical realization of those cavities, however, can be quite involved.

*Combination of resonant and non-resonant methods – Our Proposal*

The beforehand presented survey demonstrates strengths and limitations of several existent methods. We propose an experimental technique, which combines the advantages of both resonant and non-resonant methods. Our approach is based on Mie theory which describes electromagnetic scattering from the sphere [17]. Mie theory links geometry, the material parameters of a sphere, and resonant peaks in the scattering spectrum. Mie coefficients, describing the complete set of modes, have explicit closed form expressions in the form of Bessel functions. In the case of small high-index particles, Padé expansion for the so-called magnetic coefficient $b_n$ can be applied[18]. In this case, the resonant condition is given by $\varepsilon_r = -2.07 + 10.02/x^2 + 1.42x^2 - 2ix(1.06 - 0.77x^2)$, where the dimensional parameter $x = \pi D/\lambda_{res}$. $\lambda_{res}$ is the resonant wavelength, $\varepsilon_r$ is the sphere's relative permittivity, and $D$ is the sphere's diameter. This equation assumes relative permeability of the sphere $\mu_r=1$. In the case of small high-epsilon sphere, the main contribution to the resonant condition comes from the second term on the right hand side of the inline equation, written above (also the first and the third terms partially cancel each other – this was checked by substituting the range of relevant parameters and less than 1-2% deviation was found). Under those assumptions, the resonant condition for the first magnetic (dipolar) mode is given by:

$$\lambda_{res} \approx \sqrt{\varepsilon_r} \cdot D, \qquad (1)$$

where similar approximation were used in, e.g. [19]. If a broad scattering spectrum is observed, the first magnetic dipole resonance at the lowest frequency obeys the condition of Eq. 1. In this way, we obtain the real part of the permittivity, while its imaginary component can be found by analyzing the spectral widths. Both radiative and material losses, as well as coupling with antenna contribute to this value. In addition, overlapping resonances can further complicate direct retrieval. A complimentary method can be based on Kramers-Kronig relations, which link real and imaginary parts via an integral relation. However this approach has significant limitations when available experimental data has a finite bandwidth and a relatively small number of experimental points forms the spectra[20]. Here we found the first method of spectral resonance width to be more accurate, if several simplifying assumptions are made: (i) material losses prevail over radiation leakage, (ii) the spectrum has well defined and separated resonant features. In this case, the imaginary part of permittivity can be calculated via Q-factor relation:

$$\varepsilon'' = \frac{\varepsilon'}{Q}. \qquad (2)$$

Based on the formulation of Eqs. 1 and 2, we propose a new technique for parametric retrieval. Our method is calibration-free and does not require expensive additional equipment (apart from a scalar network analyzer, phase information is not mandatory). Furthermore, the data post-processing is transparent and straightforward, which makes the method to be cheap and easy to handle.

**II. METHODS**

*Liquid Solutions*

In order to retrieve permittivity within a desired frequency range, the diameter of the sphere $D$ should be varied continuously. This was implemented by using a stretchable latex balloon, which contained a liquid solution. The volume of the incompressible liquid was controlled with a calibrated injecting syringe, connected to the latex enclosure via a capillary. Schematic representation of the setup is depicted on Fig. 1, while its photograph appears in Fig. 2(a). An empirical relation between the surface tension of the balloon, the pressure, supplied by the syringe, and the overall volume of the sphere was obtained. This apparatus allowed us to achieve almost continuous control over the dimensions, which is the key factor for ensuring broadband operation of this resonance-based method. An additional important technical detail involves monitoring of the shape of the balloon, which should be kept as spherical as possible in order to employ Eq. 1 for the permittivity retrieval.

The resonance measurement was made by approaching the structure with a magnetic loop antenna (Fig. 1) positioned on top of the sphere. The antenna's feed was connected to Agilent E8362C VNA and $S_{11}$ parameter was retrieved. According to Mie theory, the first dip (the lowest in frequency) in the reflection coefficient corresponds to the magnetic dipole mode, the resonance condition of which is described by Eq. 1. In order to avoid resonance hybridization, the internal resonance of the loop antenna should be engineered to fall outside of the frequency window, where permittivity is retrieved. In this case, the loop resonance frequency was around 10 GHz, while all the resonances of spheres fall within 0.5-4 GHz range.

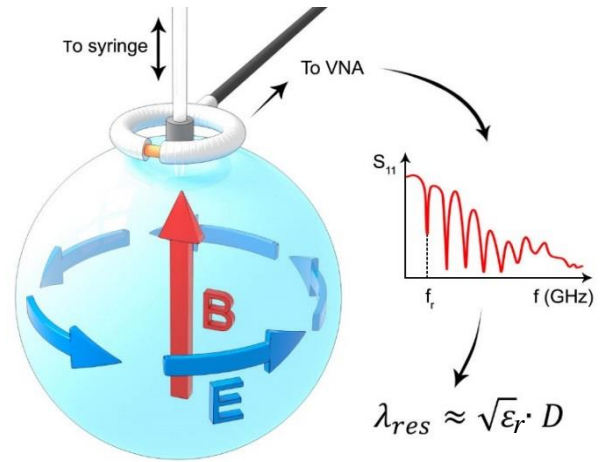

*Figure 1. Schematic representation of the experimental setup for extraction of complex permittivity. Arrows illustrate the electric (blue) and magnetic (red) vector fields, corresponding to the magnetic dipole resonance. The sphere's dimensions can be dynamically tuned, shifting the corresponding magnetic dipole resonance in frequency. A non-resonant probe (loop antenna) is connected to a vector network analyser (VNA) and placed above the sphere. Permittivity is retrieved by monitoring spectral dips in the reflection coefficient from the probe ($S_{11}$-parameter) – $S_{11}$ spectrum is schematically depicted in the right inset. Resonant frequency $f_r$, corresponding to the dipole magnetic resonance, along with the resonance quality factor, allow retrieval of complex permittivity in a broad spectral range, defined by the set of available sphere's radii.*

First, numerical simulation was performed to estimate the accuracy of the proposed method. A spherical thin shell filled with distilled water, was analyzed numerically with Frequency Domain method, implemented in CST Microwave Studio. Permittivity obtained with commercial dielectric assessment kit (DAK)[18] was used

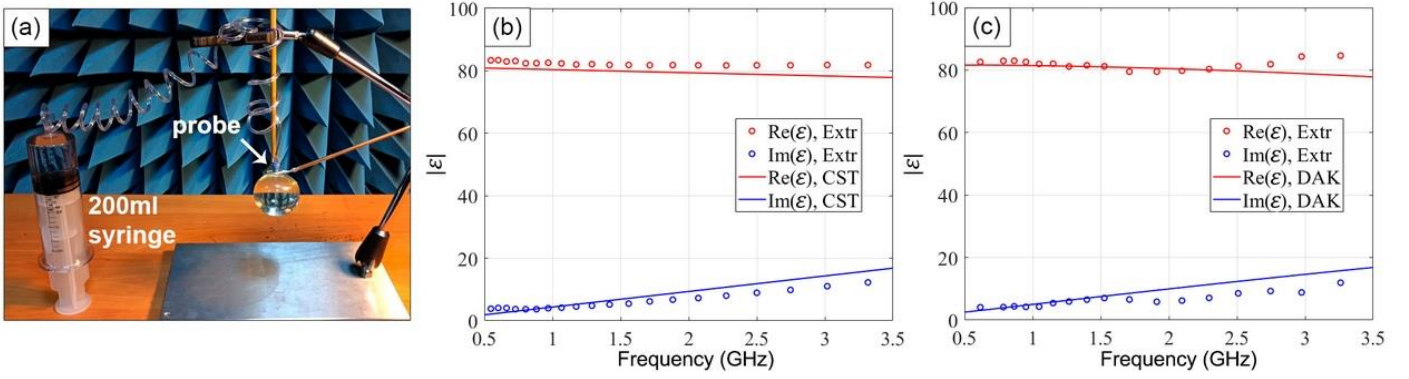

*Figure 2. (a) Photograph of the experimental setup for parametric retrieval of liquid solutions. (b) Permittivity spectrum of distilled water - real (red) and imaginary (blue) parts of permittivity. Solid lines – permittivity, extracted with a commercial dielectric assessment kit (DAK). Circles – permittivity, retrieved from numerical modeling with the subsequent post-processing via Eqs 1 and 2. Etalon permittivity (solid lines, DAK extraction) is used as an input to the modeling. Deviations between solid lines and circles demonstrate the accuracy of the method. (c) Complex permittivity of distilled water (red and blue colors stay for real and imaginary pats, correspondingly). Solid lines – values, extracted with DAK. Circles – values, extracted experimentally with setup from panel (a) by fitting resonance conditions, defined by Eqs. 1 and 2.*

as an input to the modeling. Resonant spectra for different sphere's radii were obtained and post-process with the help of Eqs. 1 and 2. Ideally, the results of this resonant type of retrieval should return the input data, while any deviations will indicate systematic errors and inaccuracies. Fig. 2(b) summarizes this investigation - complex permittivity of distilled water in the range of 0.7-5 GHz was retrieved. The liquid's temperature during the measurements was 16.3 degrees Celsius. A three-step calibration was performed, and permittivity was automatically obtained by using commercial software. DAK data was then compared to the resonant retrieval. The relative systematic inaccuracy of the method is in the range of 5% and provides a reasonable approximation of the reality. Deviation grows with the frequency as the result increase of losses. In order to perform a fair comparison and to factor out the probable influence of the liquid preparation protocol, we relied on the experimental (DAK data) rather than the tabulated data

After estimating the accuracy of the method numerically, experimental retrieval was performed. The experimental setup is shown in Fig.2 (a). The sphere was formed by a shell made of widely available medical latex and was fastened with threads at the end of a capillary. The other end of the capillary was connected to a 200 ml syringe filled with a liquid (distilled water in this case) so the sphere's radius could be varied. The loop antenna was placed around the supplying capillary, which position was found to be optimal in terms of ensuring a constant distance between the antenna and the surface of the sphere as the dimensions changed. We measured the $S_{11}$ parameter of the loop antenna with Agilent E8362C VNA for 17 different diameters of the sphere between 10 and 60 mm. The obtained data was processed using the same procedure as in the numerical modeling (Eqs. 1 and 2). The extracted values of real and imaginary parts of permittivity are shown in Fig.2 (c) (red and blue circles) together with the reference DAK measurement (solid red and blue lines). Reasonable agreement was observed, however the deviation becomes greater due to losses which increase at higher frequencies.

*Solid materials*

The proposed method can also be applied to solid materials. However, a set of samples with different dimensions should be prepared with available fabrication techniques (e.g. machine milling). Another possible and even more valuable application of the resonance method could be as a quality check of materials, where responses of new fabricated samples are compared with a reference sample. Here we demonstrate this capability by investigating six identical high-quality ceramic spheres of 15 mm diameter (Fig. 3(a)). The deviation of the radii was estimated to be in the order of 1 mm and less, according to measurements. The deviation in the resonant locations were observed to be in the range of less than 3% (Fig. 3(b) - $S_{11}$ spectra, acquired with the same loop antenna probe). Reference characteristic of the ceramics was 16 for real and 0.642 for imaginary part at 5 GHz frequency.

The calculated values of the real and imaginary parts of the spheres' permittivity are presented in the Table 1. For all the samples investigated, the losses were significantly higher than the reference data, while the real part of permittivity is quite accurate. We suggest that the losses depend on the internal granularity of ceramics obtained during the fabrication process. Fabrication methods of ceramic elements involve agglomeration of ceramic powder at high temperature and high pressure, which can result nontrivial material's microstructure.

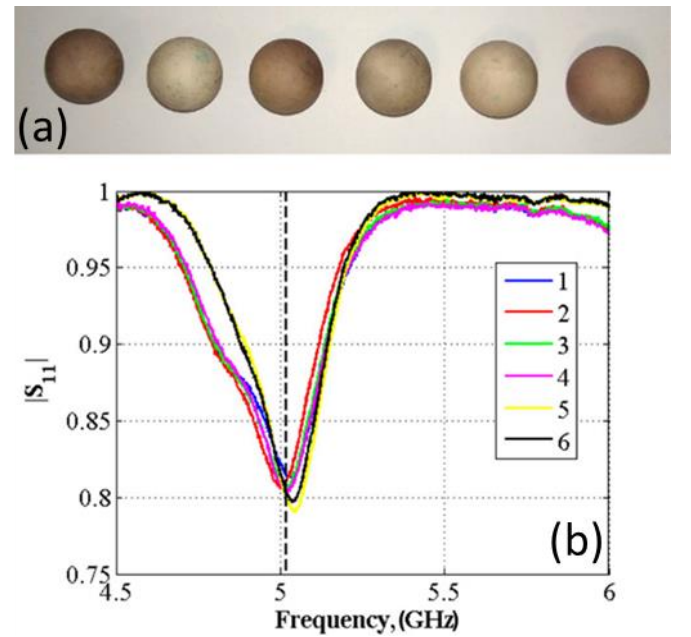

*Figure 3. (a) A photo of the ceramic samples. (b) Experimental $S_{11}$ spectra of the samples. Dashed line shows the average position of the magnetic dipole resonance frequency.*

| № | ε' | ε'' | ε' (ref) | ε'' (ref) | Δε', % | Δε'', % |
|---|-------|------|----------|-----------|--------|---------|
| 1 | 15.78 | 0.78 |          |           | 1.37   | 20.88   |
| 2 | 16.00 | 0.72 |          |           | 0%     | 12.53   |
| 3 | 15.91 | 0.73 | 16       | 0.64      | 0.56   | 14.10   |
| 4 | 15.88 | 0.68 |          |           | 0.75   | 6.32    |
| 5 | 15.73 | 0.54 |          |           | 1.69   | 15.60   |
| 6 | 15.76 | 0.56 |          |           | 1.50   | 12.54   |

*Table 1. Retrieved real and imaginary parts of permittivity of 6 identical ceramic samples (from left to right, Fig. 3(a)).*

## IV. DISCUSSION

It is important to consider the limitations of the proposed method, namely, material losses which make the resonant response less pronounced. We therefore performed a set of numerical simulations, where the real part of the permittivity was kept constant (80) and the loss tangent was varied from $10^{-4}$ to ~1. The frequency range was taken to be between 0.5 and 3.5 GHz and the radius of the sphere was 10mm Fig. 4 demonstrates the data where $|S_{11}|$ values are presented with colors, while vertical and horizontal axes are the frequency and loss tangent, respectively. Three first resonances can be clearly seen up to a point, where losses smear out the resonant behavior. This information is virtually lost for loss tangent staring from about 0.5 and higher. It can be also seen that resonance broadening affects the data already starting from $10^{-1}$-$10^{-2}$, supporting the frequency dependent deviations on Fig. 2 (losses of water grow significantly with frequency).

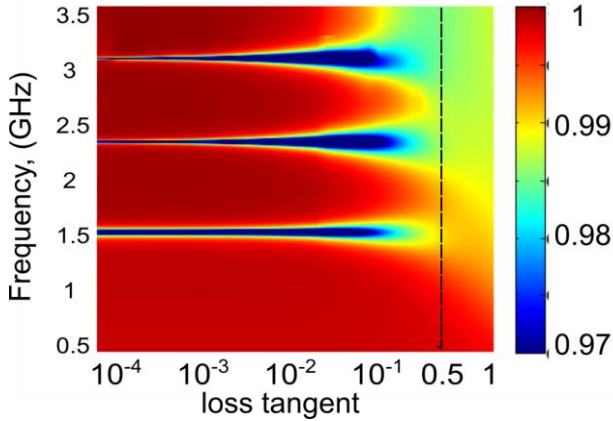

*Figure 4. Influence of losses on the resonant retrieval - $|S_{11}|$ values are presented with colors, while vertical and horizontal axes stay for frequency and loss tangent, correspondingly. Magnetic dipole resonance is the lower branch (~1.5GHz). The black dashed line shows the range of applicability (0.5 is set as the higher bound for acceptable losses). Real part of the permittivity is 80, radius of the sphere 10mm.*

## V. CONCLUSIONS

We have demonstrated a new calibration-free method for extracting complex material permittivities of liquid solutions. Our methodology combines the advantages of both the accurate resonant and broadband non-resonant approaches. Our design is based on a stretchable enclosure, which allows continuous tuning of sample's (resonant cavity) dimensions. Note, that the essence of the method is the resonant approach, while the broadband operation is achieved via continuous tuning of the resonant condition.

The key for the extraction procedure is Mie theory which links resonant conditions, dimensions, and material parameters. The proposed methodology was compared to calibrated commercially available tools and showed a good agreement within 5% accuracy over a broad frequency range (0.5-3.5GHz). The main fundamental limitation of the method is related to loss tangent, which cannot access the value of 0.5. Practical difficulties of the implementation are related to mechanical stabilization of samples and accurate maintenance of their spherical shape. Thermal control was found to be also an important factor. The temperature and chemical resistance of the enclosing spherical shell can be increased, if PVC vinyl, neoprene or nitrile are used instead of latex. In this case, thicker enclosures might be employed. As the result, a core-shell geometry (shell – the enclosure; core – a liquid under investigation) will be involved. Core-shell geometries can be also analyzed with Mie solutions[21], which, however, become cumbersome in this case. Furthermore, the usage of thick shells might require a calibration step to account for their response. In the case of latex enclosure, shells' influence can be safely neglected.

The accuracy of the methodology can be further improved if the entire Mie resonance spectra are post-processed and complexes-valued S-parameters are analyzed. However, higher-order modes are more confined to the structure and, as the result, are more sensitive to surface imperfections and internal material losses, hence, they were excluded from the consideration in this work.

The main advantages of the current method are its straightforward implementation, clear relation between measured data and the physical origin, and cheap experimental equipment.

In terms of future practical application, this new broadband resonant approach can be employed for investigation of different types of materials, such as liquids, powders and solids. Continuous shape variation of the cavity configuration allows permittivity retrieval. Additive manufacturing techniques and selective surface metallization can allow creating complex geometries [6,22] and, as the result, can allow more accurate parametric retrieval on expense of computational and post-processing simplicity.

## Acknowledgments

The work was supported, in part, by ERC StG 'In Motion' (802279), Pazy Foundation, and Ministry of Science and Technology (project "Integrated 2D & 3D Functional Printing of Batteries with Metamaterials and Antennas").